\documentstyle[twoside,fleqn,espcrc2,epsfig]{article}

\newcommand{\AmS}{{\protect\the\textfont2
  A\kern-.1667em\lower.5ex\hbox{M}\kern-.125emS}}

\title{Probing the extremes of Seyfert activity: BeppoSAX observations of 
Narrow-Line Seyfert 1 galaxies}

\author{A. Comastri\address{Osservatorio Astronomico di Bologna,
        via Zamboni 33, I-40126 Bologna, Italy},
        \thanks{This work has received partial financial support from the 
                 Italian Space Agency (ASI contract ARS-96-70).} 
        W.N. Brandt\address{Dept. of Astr. and Astroph., Penn
          State University, 525 Davey Lab, University Park, PA 16802, USA},
        K.M. Leighly\address{Columbia Astroph. Lab., Columbia University, 
          538 West 120th Street, New York, NY 10027, USA},
        F. Fiore\address{Osservatorio Astronomico di Roma, via 
       dell' Osservatorio, I-00040 Monteporzio--Catone, Italy}, 
        M. Guainazzi\address{SAX--Science Data Center, Nuova Telespazio, 
       via Corcolle 19, I-00131 Roma, Italy},
        G. Matt\address{Dip. di Fisica, Universit\`a degli Studi
          ``Roma Tre", Via della Vasca Navale 84, I--00146 Roma, Italy},
        G.M. Stirpe$^{\rm a}$}
       
\begin{document}

\begin{abstract}
Results are presented for the first year of observations 
of a selected sample of Narrow-Line Seyfert 1 galaxies (Ton~S~180, 
RE J1034+396, Ark 564) obtained 
with the imaging instruments onboard BeppoSAX.
These are the first simultaneous broad band (0.1--10 keV) spectra
so far obtained for this class of objects.

\end{abstract}

\maketitle

\section{INTRODUCTION}

Narrow--line Seyfert 1 galaxies (NLS1) are defined by their optical 
emission line properties. They lie at the lower end of the broad line 
width distribution for the Seyfert~1 class
with typical values of the H$\beta$ FWHM in the range 500--2000
km $s^{-1}$. The
[O~{\sc iii}]/H$\beta$ ratio is $<$ 3, and
strong  Fe~{\sc ii} and high ionization iron lines are common among NLS1.
They are characterized by 
unusually strong soft X--ray emission with very steep ($\Gamma \simeq$ 3--5)
spectral slopes in the ROSAT PSPC band \cite{BBF96}.
Large-amplitude and rapid X-ray variability is
common among NLS1. 
The variability amplitude is significantly larger than that found
in ``normal" Seyfert galaxies 
\cite{Karen97}. Moreover, there is
evidence for giant-amplitude X-ray variability
(up to about two orders of magnitude) in a few objects
\cite{Bol97,FH96,Brandt95,Gru95a,Gru95b}. It is notable
that such extreme variability properties have been discovered,
so far, only in NLS1.
In the harder $\approx$ 2--10~keV
energy band, recent ASCA observations
have shown that NLS1 can have very different behaviour from
classical broad-lined Seyfert~1s and quasars.
A comparative ASCA study of a
large sample of NLS1 and broad-line Seyfert~1s
revealed that the $\approx$~2--10~keV
ASCA spectral slopes of NLS1 are
generally steeper than those of broad line objects at
a high statistical confidence level \cite{Brandt97}.

The optical and X--ray properties of ultrasoft NLS1 
have been discussed in detail by \cite{BB97}
adopting a principal component analysis technique \cite{BG92}. 
The results indicate
that they tend to lie at the ends of distributions of Seyfert quantities 
suggesting that they may have extremal values of some important
physical parameter.

In order to increase the statistics of hard X--ray data
for NLS1 galaxies and to better investigate their underlying 
physical processes we have undertaken a program of
BeppoSAX observations of a sizeable sample
of NLS1. We have selected, for the first
year of observations, some of the brightest and
most variable NLS1 previously observed by ROSAT and/or ASCA.
The spectral capabilities of the detectors onboard BeppoSAX
and especially the relatively large effective area at high energy ($>$ 2 keV)
will allow a better study of the high energy properties of NLS1.
In this paper we present the first simultaneous 0.1--10 keV spectral and
variability properties for 3 objects, namely: Ton S 180, RE J 1034+396 and
Ark 564.

\section{TON S 180}

Ton~S~180, an optically bright galaxy with an optical spectrum typical 
of NLS1, is a strong and variable extreme ultraviolet source detected by the 
EUVE satellite \cite{Vennes95}. Not surprisingly it is also 
bright in the ROSAT PSPC energy range. Amplitude
variations up to a factor 2 on timescales ranging from hours to days 
have been observed. The 0.1--2.4 keV spectrum is steep with photon slopes 
ranging between 2.7 and 3.2 \cite{Fink96}.
A relatively short ($\sim$ 20 ksec) BeppoSAX exposure was performed 
on 1996 December 3. A detailed discussion of this observation can be found 
elsewhere \cite{Comastri98}. 
The most relevant results can be summarized as follows.
Relatively rapid flux 
variations, by a factor 2 on a timescale of a few 10$^4$ s,
in both the soft (0.1--3.0 keV) and hard (3--10 keV) X--ray band are clearly
present (first two panels of Fig.~1), without any significant evidence
of spectral variability (third panel of Fig.~1).

\begin{figure}
\epsfig{file=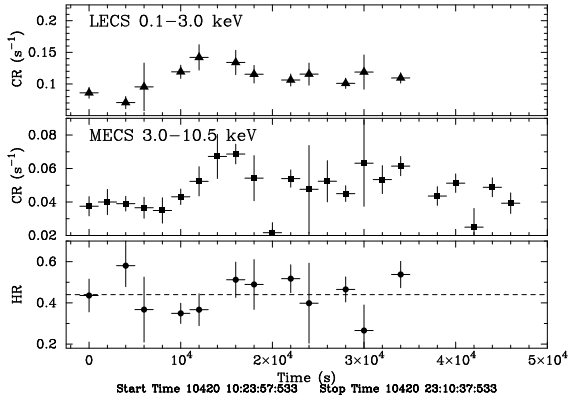,  height=8.cm, angle=-90, silent=}
\caption{BeppoSAX light curves for Ton S 180 in the soft band (top), 
medium energy band (middle) and hardness ratio light curve (bottom).}
\end{figure}

The 0.1--10~keV spectrum requires at least two components: (1) a steep
$\Gamma \simeq$ 2.7 strong soft component below 2~keV which contains a
large fraction of the
overall energy output, and (2) a weak hard tail with a 2--10~keV slope
$\Gamma \simeq$ 2.3
that is significantly steeper than the average value found in
``normal'' Seyfert~1s and quasars. 
There is evidence for iron line emission at $\sim$ 7~keV (Fig.~2). 
When fitted with
a simple Gaussian, the best fit centroid indicates a high (H--like) 
ionization state. Ionized iron line emission has been also found in the
ASCA observation (Leighly et al. in preparation).

\begin{figure}
\epsfig{file=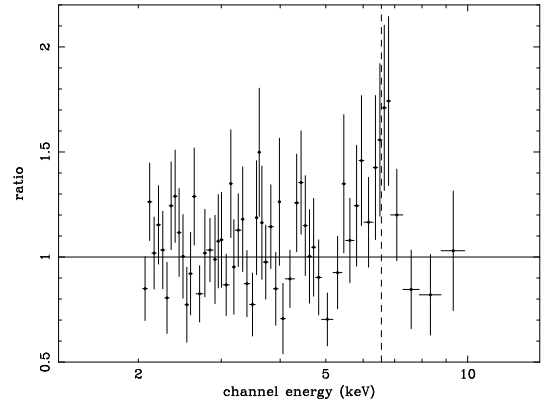,  height=8.cm, angle=-90, silent=}
\caption{Residuals from a single power law fit ($\Gamma = 2.21\pm0.12$) 
in the 2--10~keV band for Ton S 180. 
The vertical dotted line is at the energy of H--like iron (6.97 keV).
The energy scale is in the rest system of the observer.}
\end{figure}

\section{RE J 1034+396}

RE J 1034+396 belongs to the small sample of AGNs detected 
by the Wide Field Camera onboard ROSAT at very soft (below 0.18 keV)
X--ray energies. The PSPC spectrum is extremely soft 
and can be modelled in terms of two black bodies with temperatures
of $\sim$ 40 and $\sim$ 100 eV plus an extremely weak hard power law
component. Follow--up ASCA observations revealed that the power law
was unusually steep: $\Gamma \simeq$ 2.6. 
An ionized iron line at $\sim$ 6.7 keV is marginally (at 90\% confidence)
required \cite{PDO95}.

The source was clearly detected by BeppoSAX on 1997 April 18
in both LECS and MECS with a $\sim$ 40 ksec exposure. 
There is no evidence of flux variability in the BeppoSAX observation, moreover 
the flux level is in good agreement with the previous ASCA value suggesting
a remarkably constant X--ray flux over several years.
The 0.1--10 keV spectral energy distribution is dominated 
by a strong soft component below $\sim$ 2 keV which accounts for more than
90\% of the total luminosity, while at higher energies the 
source is much weaker. 
The RE J1034+396 continuum emission is best fitted by two blackbodies 
with temperatures of 55$\pm$10 eV and 155$\pm$20 eV plus a steep power law
$\Gamma=2.20\pm0.24$ (Fig.~3). The overall spectral shape is in good 
agreement 
with previous results \cite{PDO95} even if the values of the spectral 
parameters are slightly different. There is no evidence of iron line emission
but it should be noted that the signal to noise at high energies is rather poor.

\begin{figure}
\epsfig{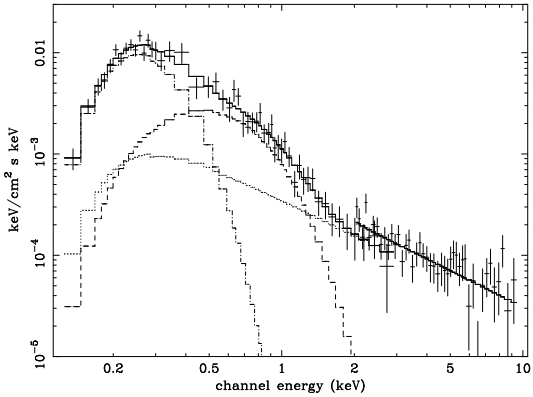}
\caption{The best fit unfolded spectrum of RE J1034+396, consisting
of two blackbodies for the low energy emission plus a power law
(see text for details).}
\end{figure}

On the basis of optical and HST ultraviolet observations, 
it has been argued \cite{PMS98} that RE J1034+396
has one of the hottest big blue bumps of any Seyfert. It is completely
shifted out of the UV band and is thus visible at Extreme UV and X--ray 
wavelengths. This would be consistent with ASCA and BeppoSAX spectral fits.
The two blackbodies can be considered as a first order approximation of the
high energy tail of a hot optically thick accretion disk.

\section{ARK 564}

Arakelian 564 is a bright NLS1 observed by ROSAT \cite{BBF96}.
The 0.1--2.4 keV PSPC spectrum appears to be complex as a simple
steep power law ($\Gamma \simeq 3.4$) provides a very poor fit \cite{B94}.
The residuals of the power law fit suggest the presence of spectral
features typical of warm absorbing gas. 
Either a line at $\sim$ 0.8 keV or an edge at $\sim$ 1.15 keV
improve significantly the fit quality, 
however the PSPC spectral resolution prevented detailed
analysis.

A $\sim$ 20 ksec BeppoSAX observation has been recently 
performed (1997 November 14). 
The first preliminary results of
the BeppoSAX observation are summarized below, while a more detailed
analysis will be deferred to a future paper.

\begin{figure}
\epsfig{file=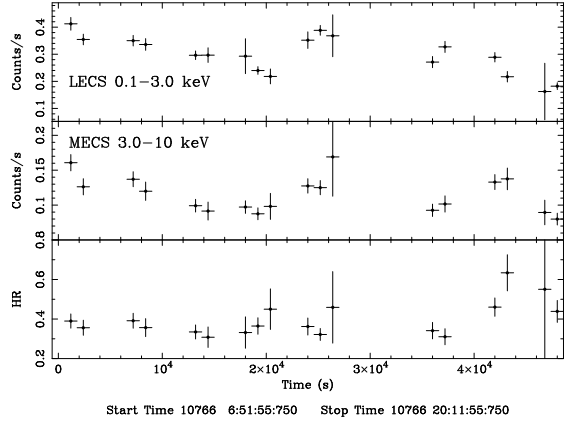, height=8.5cm, angle=-90, silent=}
\caption{BeppoSAX light curves for Ark 564 in the 0.1-3.0 keV LECS band (top) 
3.0-10 keV MECS energy range (middle) and hardness ratio light curve (bottom).}
\end{figure}

The light curves in the soft 0.1-3.0 keV 
LECS energy range and in the 3.0-10 keV MECS energy range are shown
in the first two panels of Fig.~4. 
High amplitude variability, with a doubling time of the order of 
10$^4$ s has been clearly detected at all energies.
There is also some evidence of rapid variability at the end of the observation
possibly associated with spectral changes as suggested by the hardness ratio
light curve (bottom panel of Fig.~4). 

The overall 0.1--10 keV spectrum requires at least two components.
Either a broken power law or a blackbody plus a power law provide 
a relatively good description of the continuum emission.
In both cases the 0.1--2.0 keV flux is about one order of magnitude greater
than the 2--10 keV flux and the 2--10 keV power law is very steep
($\Gamma \simeq 2.4$). Iron line emission at $\sim$ 6 keV
is clearly present (Fig.~5).

\begin{figure}
\epsfig{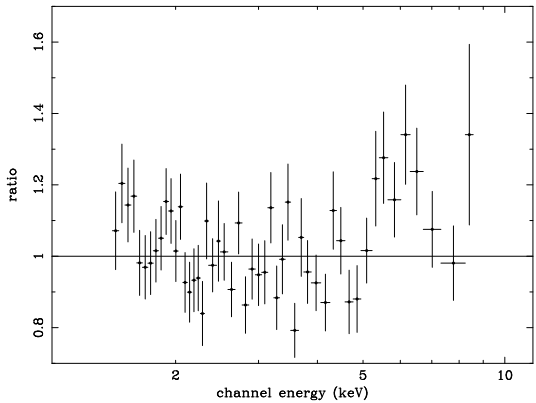}
\caption{Residuals from a single power law fit ($\Gamma = 2.38\pm0.07$) 
in the 2--10~keV band for Ark 564. 
The energy scale is in the rest system of the observer.}
\end{figure}

The best fit energy: $6.22\pm0.30$ keV
indicates neutral or mildly ionized iron. The line appears to be 
broad ($\sigma = 0.65$ keV), however its
width is only weakly constrained by the present data.

\begin{figure}
\epsfig{file=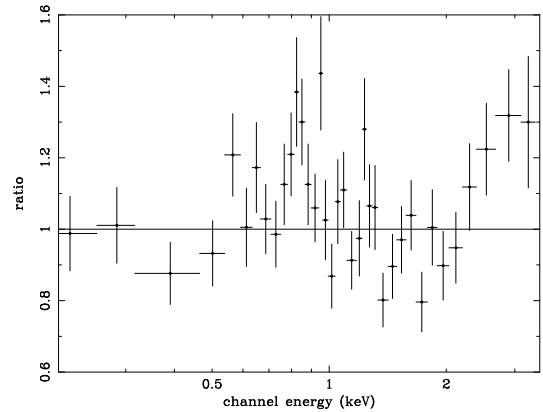, height=8.cm, angle=-90, silent=}
\caption{Residuals from a single power law fit ($\Gamma = 2.90\pm0.08$) 
in the 0.1--3~keV band for Ark 564. A line-like feature at $\sim$ 0.8 keV
and edge--like structures in the 1--2 keV band are present.}
\end{figure}

A single power law gives a poor fit to the low energy LECS spectrum
(Fig.~6).
The most prominent deviations from such a model are 
suggestive of the presence of ionized absorbing gas 
which is common among Seyfert galaxies \cite{R97}.
When a narrow Gaussian line is added a significant improvement
is obtained  
($\Delta\chi^2 \simeq 40$). The best fit line energy and equivalent width are:
$E = 0.86 \pm 0.04$ keV and 55$\pm$24 eV respectively.
An even better description of the low energy data 
($\Delta\chi^2 \simeq 15$ with respect to the power law plus line)
was obtained by adding an absorption edge at $E = 1.38^{+0.17}_{-0.09}$
keV with depth $\tau = 0.30\pm0.12$. 
A detailed discussion on the origin of these spectral features is 
beyond the scope of this paper, however 
we note that these results are somewhat surprising as 
the detected spectral features are rather different from those found
in classical Seyfert 1 galaxies \cite{R97}. Moreover
there are no obvious lines and edges expected at these energies 
as discussed in \cite{B94,L97}. Finally it should also be noted that
fitting the soft X--ray spectrum with a blackbody model the line 
and edge are still present although at a lower level of significance.

\section{DISCUSSION}

The X--ray spectral properties of the 3 NLS1 observed by BeppoSAX 
are remarkable when compared with those of classical Seyfert 1s and
quasars. 
The spectral shapes are summarized in Table 1. The broken power law
fits are to be considered as a first approximation of the 
overall 0.1--10 keV X--ray spectrum. 
The soft X--ray spectra of RE J1034+396 and Ark 564 can also be fit with the
high energy tail of thermal emission models.
A two--component model provides an adequate description of the 0.1--10 keV 
continuum for all the objects in analogy with several broad lined AGNs. 
However, the slopes of the two components are rather different from 
those in broad line AGNs \cite{Ciccio92}.
The soft component is much stronger than in ``normal" Seyfert 1s, while
the 2--10 keV power law slopes are much steeper than the average 
value ($\Gamma \simeq 1.9$) for Seyfert 1s and
quasars in the same energy range \cite{N97,Ree97}.
The large relative intensity of the soft component and the steep 2--10 keV
slopes (see also \cite{Brandt97}) have important consequences.
Even if the observation of iron lines suggests that reprocessing is 
occuring at some level, it appears likely that the strong soft component 
cannot be due only to disk reprocessing, unless there is highly anisotropic
emission or a high energy spectrum extending up to extremely high energies.

\begin{table*}[hbt]
\setlength{\tabcolsep}{1.5pc}
\newlength{\digitwidth} \settowidth{\digitwidth}{\rm 0}
\catcode`?=\active \def?{\kern\digitwidth}
\caption{Summary of the BeppoSAX spectral fits}
\label{tab:effluents}
\begin{tabular*}{\textwidth}{@{}l@{\extracolsep{\fill}}lccccc}
\hline
                 & \multicolumn{2}{c}{Single power law (2-10 keV)} 
                 & \multicolumn{3}{c}{Broken power law (0.1-10 keV)} \\
\cline{2-6}
                 & \multicolumn{1}{c}{$\Gamma$} 
                 & \multicolumn{1}{c}{Flux}
                 & \multicolumn{1}{c}{$\Gamma_s$}
                 & \multicolumn{1}{c}{$E_{break}$}
                 & \multicolumn{1}{c}{$\Gamma_h$}    \\
\hline
TON S 180  ~~~      &  $2.21\pm0.12$ & $4.2\cdot10^{-12}$ & $2.68\pm0.07$ & 
                  $2.5\pm0.7$ & $2.29\pm0.15$ \\
RE J1034   ~~~      &   $2.20\pm0.23$ & $9.3\cdot10^{-13}$ & $3.24\pm0.06$ & 
                   $2.5\pm0.6$ & $2.08\pm0.35$ \\
ARK 564    ~~~      & $2.38\pm0.07$ & $1.5\cdot10^{-11}$ & $2.75\pm0.05$ & 
                   $2.2\pm0.5$ & $2.37\pm0.08$  \\
\hline
\multicolumn{6}{@{}p{120mm}}{The absorption column density has been fixed at 
the Galactic value in all the fits. Fluxes are in erg cm$^{-2}$ s$^{-1}$,
break energies in keV. Errors are at 90\% confidence for one interesting 
parameter.}
\end{tabular*}
\end{table*}

\begin{figure}
\epsfig{file=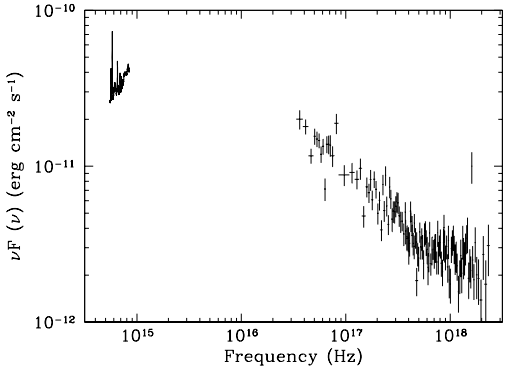, height=7.cm, angle=-90, silent=}
\caption{The broad band energy distribution of Ton S 180
from optical to X-ray energies \cite{Comastri98}.}
\end{figure}

\begin{figure}
\epsfig{file=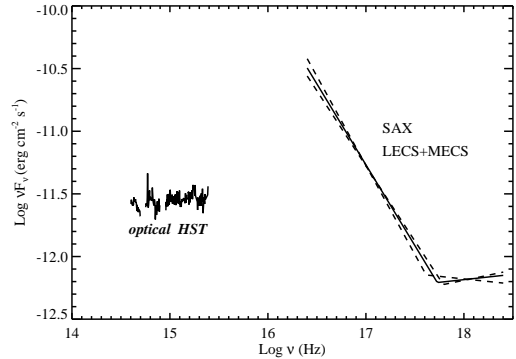, height=5.cm, angle=0, silent=}
\caption{The broad band energy distribution of RE J1034+396
from optical to X-ray energies \cite{PMS98}.}
\end{figure}

The strong soft component could lead to a strong Compton cooling of the hot 
corona electrons and so to a steep hard tail if a significant fraction of 
the gravitational energy is dissipated in the disk phase \cite{HM93}.
This hypothesis is also supported by the similarities between the NLS1
X--ray spectra and those of Galactic black hole candidates (GBHC) in their 
high state as first suggested in \cite{BBF96,PDO95}.
The high states of GBHC are thought to be triggered by increases in the 
accretion rate resulting in strong thermal emission from a disk accreting near
the Eddington limit. If this is the case the disk surface is expected 
to be strongly ionized which fits nicely with the detection of a ionized 
line in Ton~S~180 \cite{Comastri98}.
A high $L/L_{Edd}$ ratio would also be consistent with the narrowness of
the optical lines in NLS1. In fact 
the optical line width is inversely proportional to $L/L_{Edd}$ if the 
broad line region is virialized and its radius is a function
of luminosity alone \cite{BBF96,LF97,WB98}.
The lack of iron line emission in RE J 1034+396 and the low iron ionization
state in Ark 564 raise some problems with this interpretation. 

Despite the similar X--ray spectral energy distributions (See Table 1) the  
NLS1 observed so far differ in several other properties.
Ton~S~180 and Ark~564 show relatively rapid X--ray variability on timescales
of the order of 10$^4$ s, and  spectral variability is probably present 
in Ark 564. On the other hand the RE J1034+396 
light curve appears to be constant over several years.

The imprints of highly ionized absorbing gas are visible in the
soft X--ray spectrum of Ark~564, while Ton~S~180 and RE J1034+396 are 
characterized by relatively smooth soft X--ray spectra.

It is also worth noting that RE J1034+396 has strong high ionization iron lines
but weak Fe~{\sc ii}, while the other two objects have 
strong Fe~{\sc ii} \cite{Comastri98}
emission and not so strong high ionization iron lines.

Finally the spectral energy distributions (SED) from optical to X--rays
are also different.
The SED of Ton~S~180 and RE J1034+396 are displayed in  
Figures 7 and 8 respectively.

The data for Ton~S~180 suggest that the energy density peaks somewhere
in the ultraviolet. The optical--UV light dominates the energy output 
with a behaviour similar to that of other quasars \cite{Z97}.
The situation is reversed for RE J1034+396 where most of the
energy density is found at soft X--ray wavelengths, while the UV continuum
is rather weak \cite{PMS98}.
 
Further BeppoSAX observations and coordinated optical campaigns 
are clearly needed to clarify some of these issues.

\end{document}